\documentclass[10pt,aps,prd,twocolumn,floats,floatfix,showpacs,superscriptaddress,nofootinbib]{revtex4-2}
\usepackage[utf8]{inputenc} 

\usepackage{graphicx,mathtools,amssymb,amsmath,amsthm,amsfonts,epsfig,epsf,bm,float,subfig,multirow}
\usepackage[outdir=./]{epstopdf}
\usepackage[linktocpage]{hyperref}
\hypersetup{hidelinks}
\hypersetup{pdfstartview=}
\hypersetup{
    colorlinks=true,
    linkcolor=blue,
    filecolor=magenta,      
    urlcolor=BrickRed,
    citecolor=BrickRed,
}
\usepackage[usenames]{color}
\usepackage{tensor}
\usepackage{csquotes}
\usepackage{mathrsfs}
\usepackage{autobreak}
\usepackage{physics}
\usepackage{cancel}

\usepackage[font=small,labelfont=bf, justification=Justified,
   format=plain]{caption}

\usepackage{slashed}

\def\nn{\nonumber}
\def\del{\partial}
\def\L{\mathcal{L}}
\def\be{\begin{equation}}
\def\ee{\end{equation}}
\def\bea{\begin{eqnarray}}
\def\eea{\end{eqnarray}}
\usepackage{pifont}

\usepackage[dvipsnames]{xcolor}

\begin{document}

\title{Horndeski speed tests with scalar-photon couplings}

\author{\textbf{Eugeny Babichev}}
\email{babichev@ijclab.in2p3.fr}
\affiliation{Universit\'e Paris-Saclay, CNRS/IN2P3, IJCLab, 91405 Orsay, France}

\author{\textbf{Christos Charmousis}}
\email{christos.charmousis@ijclab.in2p3.fr}
\affiliation{Universit\'e Paris-Saclay, CNRS/IN2P3, IJCLab, 91405 Orsay, France}

\author{\textbf{Benjamin Muntz}}
\email{Benjamin.Muntz@nottingham.ac.uk}
\affiliation{Nottingham Centre of Gravity, University of Nottingham,
University Park, Nottingham NG7 2RD, United Kingdom}
\affiliation{School of Physics and Astronomy, University of Nottingham, University Park, Nottingham NG7 2RD, United Kingdom}

\author{\textbf{Antonio Padilla}}
\email{Antonio.Padilla@nottingham.ac.uk}
\affiliation{Nottingham Centre of Gravity, University of Nottingham,
University Park, Nottingham NG7 2RD, United Kingdom}
\affiliation{School of Physics and Astronomy, University of Nottingham, University Park, Nottingham NG7 2RD, United Kingdom}

\author{\textbf{Ippocratis D. Saltas}}
\email{saltas@fzu.cz}
\affiliation{CEICO, Institute of Physics of the Czech Academy of Sciences, Na Slovance 2, 182 21 Praha 8, Czechia}

%%%%

\begin{abstract}
We revisit multi-messenger constraints from neutron star mergers on the speed of propagation of gravitational and electromagnetic waves in Horndeski and beyond Horndeski theories. By considering non-trivial couplings between the dark energy field and the electromagnetic sector, the electromagnetic wave can propagate through the cosmological background at non-unit speed, altering the phenomenological constraints  on its gravitational counterpart. In particular, we show  that recent models derived from a Kaluza-Klein compactification of higher dimensional Horndeski models fall into a broader class of theories disformally related to those whose gravitational waves propagate with unit speed. This disformal equivalence can, however, be broken by the gravitational couplings to other sectors with interesting phenomenological consequences. We also consider higher order couplings between the scalar and the photon with second order field equations, and show that they are not compatible with constraints coming from multi-messenger speed tests and the decay of the gravitational wave. 
\end{abstract}

\maketitle

%%%%%%%%%%%%%%%%%%%%%%%%%
\section{Introduction}\label{sec:Introduction}
%%%%%%%%%%%%%%%%%%%%%%%%%
 Neutron star mergers are a powerful multi-messenger probe of General Relativity and its extensions \cite{Clifton:2011jh,Joyce:2014kja,Heisenberg:2018vsk}.  Famously, the merger of two neutron stars at a redshift of $z \sim 0.01$ resulted in the LIGO/Virgo detection of gravitational wave GW170817 alongside the observation of its optical counterpart, GRB170817A  \cite{LIGOScientific:2017vwq,Goldstein:2017mmi,Savchenko:2017ffs,LIGOScientific:2017zic,LIGOScientific:2017ync}. This single event placed extremely strong constraints on deviations from General Relativity in the dynamics of the late universe \cite{Copeland:2006wr}.  This is because interactions between the massless spin 2 field and additional light degrees of freedom can cause the gravitational wave to  propagate through the cosmological medium at a speed different to the electromagnetic wave \cite{BeltranJimenez:2015sgd,Lombriser:2015sxa,Lombriser:2016yzn}.  This is problematic because the data received from neutron star mergers that suggests the two waves are propagating at the same speed to an accuracy of one part in $10^{15}$.  As a result,  multi-messenger events have been able to rule out large classes of modified gravity models at late times \cite{Sakstein:2017xjx,Ezquiaga:2017ekz,Creminelli:2017sry,Baker:2017hug,Amendola:2017orw,Langlois,Marco,boss,Kase}. 

From within the Horndeski \cite{horn,george} and beyond Horndeski \cite{bh1,bh2} family of scalar-tensor models, only a handful survived the cull, including models with conformal scalar couplings such as Brans-Dicke gravity \cite{bd},  and so-called Kinetic Gravity Braiding models \cite{kgb}. That said, these conclusions are subject to several caveats. In particular, in \cite{Copeland:2018yuh} it was shown that the constraints were weakened when you include knowledge of the equations of motion for the cosmological  background, although these loopholes were later closed by considering inhomogeneities \cite{Bordin:2020fww}. A related idea saw the speed of the gravitational wave driven dynamically to unity through a non-minimal coupling between dark energy and dark matter \cite{betoni}. In \cite{scott}, it was observed that the frequency scales of the neutron star event lie close to the strong coupling scale of a large class of dark energy models raising legitimate doubts as to the validity of the constraints in models that  exhibit Vainshtein screening \cite{vain1,vain2,vain3}. For a discussion of strong coupling concerns and ultra-violet effects in these models, see \cite{amps,unit, ippo, tolley, oranges}.

Recently, another loophole was identified in \cite{Mironov:2024idn}, when the electromagnetic field  is generated from  generic Kaluza-Klein compactification. This introduces non-trivial couplings between the photon and the scalar that automatically line up with the graviton-scalar couplings, ensuring that the gravitational and electromagnetic waves propagate at the same speed, even in the cosmological medium. Although the waves do not propagate at unit speed, the ratio of their speeds is always one in any cosmological background, ensuring compatibility with the multi-messenger speed tests described above.

This result prompts the question: to what extent can non-trivial couplings between the scalar and the photon relax the constraints from GW170817?  The Kaluza-Klein compactifications explored in   \cite{Mironov:2024idn} immediately motivate a leading order Lagrangian density for the electromagnetic field of the form
\begin{equation}
\mathcal{L}_\gamma=    - \frac{1}{4}\left[\lambda(\varphi,X) F_{\mu\nu}F^{\mu\nu} + \mu(\varphi,X)F^{\mu}{}_{\alpha}{F}^{\nu\alpha}\varphi_\mu\varphi_\nu\right] \label{Lgam}
\end{equation}
where we denote the covariant derivative $\nabla_\mu$with a subscript, $\varphi_\mu \equiv \nabla_\mu \varphi$,  $\varphi_{\mu \nu} \equiv \nabla_\nu \nabla_\mu \varphi$ etc. and ${X \equiv g^{\mu\nu}\varphi_\mu\varphi_\nu}$. In \cite{Mironov:2024idn} the functions $\lambda(\varphi,X)$ and $\mu(\varphi,X)$ take a particular form in terms of the Horndeski potentials and also contain an overall dilaton factor. In this paper, we consider general scalar-vector-tensor Lagrangians alongside a gravitational sector described by the full set of (beyond) Horndeski Lagrangians.  Indeed, for an electromagnetic sector described by \eqref{Lgam}, we find that the constraints from multi-messenger speed tests are relaxed, as they were for \cite{Mironov:2024idn}. Although this would seem to open up the parameter space of allowed theories considerably, we show that this is not really the case.  The extended parameter space is readily obtained from a disformal transformation of the original parameter space of beyond Horndeski theories with unit speed for the gravitational wave and a minimally coupled photon, originally identified in \cite{Sakstein:2017xjx,Ezquiaga:2017ekz,Creminelli:2017sry,Baker:2017hug,Amendola:2017orw,Langlois,Marco,boss,Kase}. 

Of course, scalar-photon couplings of the form \eqref{Lgam} are not the most general: even at leading order in the electromagnetic field, other more exotic couplings are allowed that still preserve second order field equations \cite{Heisenberg:2018acv,Heisenberg:2018mxx,Kase:2018nwt}. These additional couplings are particularly interesting as they {\it cannot} be generated from a minimally coupled electromagnetic field and  a disformal transformation.  When we consider multi-messenger speed tests for this more general class of theories, we might hope that the parameter space of allowed theories is significantly extended and in a non-trivial way. However, this is not the case. Additional constraints  on the gravitational sector are obtained from the decay of gravitational waves into scalars \cite{decay}.  These are very restrictive and shut down any hope for new classes of allowed theories, beyond a disformal transformation of those originally identified in \cite{Sakstein:2017xjx,Ezquiaga:2017ekz,Creminelli:2017sry,Baker:2017hug,Amendola:2017orw,Langlois,Marco,boss,Kase}.

Although our disformally related theories are equivalent at the level of the gravitational and electromagnetic sectors, this equivalence is broken if any other sectors are minimally coupled to a different disformal combination. In this sense,  we may consider this disformal trick as a way of rescuing  some modified gravity models that were previously thought to be ruled out by multi-messenger speed tests, including some that may have interesting applications to dark energy \cite{Clifton:2011jh,Joyce:2014kja,Heisenberg:2018vsk,Copeland:2006wr} and the cosmological constant problem \cite{PadRev,BernardoRev}. Of course, in the spirit of \cite{Brax_2014,Brax:2015hma,Brax:2016did}, any non-trivial couplings between the dark energy scalar and other matter fields are subject to other constraints in both cosmology and particle physics that ought to be explicitly explored.  The most well-studied non-minimal couplings between scalars and Standard Model matter have been conformal ones. Such couplings typically predict a time variation of fundamental constants such as particle masses, or the fine-structure constant and have been constrained through the CMB, supernovae, pulsars and other observations (see e.g. \cite{Olive_2008, Arvanitaki_2015, Kaplan_2022, Bouley_2023, sakstein2023, vandeBruck2013}). Crucially, conformal couplings leave light cones unaffected, i.e propagation speeds do not change. Disformal transformations on the other hand, do not respect light-cone structure, leading to propagation speed modifications. Below we will show that it is possible to change the light-cone structure of photons and gravitons in exactly the same way, making them propagate with equal speeds. We will also comment on the phenomenological implications.

The rest of this paper is organised as follows. In the next section, we warm up our analysis by comparing the speed of propagation of the electromagnetic and gravitational waves in theories inspired by the Kaluza-Klein analysis of \cite{Mironov:2024idn}. We then show that the extended parameter space of allowed theories is really an artefact of the disformal transformation. In section \ref{sec:gen-couplings}, we consider the complete class of scalar-photon Lagrangians compatible with second order field equations, at leading order in the electromagnetic field.  Further imposing constraints on the decay of gravitational waves into scalars,  we now find no further  extension of the parameter space of allowed theories. In section \ref{sec:conc}, we conclude and comment on some phenomenological implications. 

\section{Scalar-photon couplings inspired by Kaluza-Klein compactifications}
In \cite{Mironov:2024idn}, the authors considered a five-dimensional Horndeski theory including three scalar potentials $G_2(\varphi, X), G_3(\varphi, X)$ and $G_4(\varphi, X)$,  compactified on a circle. The result was an effective four-dimensional theory of a metric $g_{\mu\nu}$, Horndeski scalar $\varphi$, $U(1)$ gauge field $A_\mu$ and a dilaton $\phi$. Identifying the $U(1)$ field with the photon,  this effective theory has its gravitational wave propagating at the same speed as the electromagnetic wave.  This effect is easily understood and should apply to any time independent compactification. The point is that the electromagnetic wave and the gravitational wave  in the four dimensional theory are just different components of the same  higher dimensional gravitational wave. In the simple example studied in  \cite{Mironov:2024idn}, the Lagrangian for the electromagnetic sector is given by 
\begin{equation}
\mathcal{L}_\gamma=    - \frac{1}{4}\phi^3\left[G_4F_{\mu\nu}F^{\mu\nu} -4 G_{4, X}F^{\mu}{}_{\alpha}{F}^{\nu\alpha}\varphi_\mu\varphi_\nu\right], \label{LgamKK}
\end{equation}
where comma denotes partial differentiation, such that $G_{4, X}\equiv \frac{\partial G_4}{\partial X}$. This motivates us to consider a generalised set-up described by the following action, 
\begin{equation}
    S= S_\text{BH}[g_{\mu\nu},\varphi]+\int \dd^4 x\sqrt{-g} \L_\gamma,
\end{equation}
where $\L_\gamma$ is given by \eqref{Lgam} and $S_\text{BH}=\int d^4 x  \sqrt{-g}\sum_{n=2}^5 {\cal L}_n$ corresponds to a scalar-tensor sector including Horndeski \cite{horn,george} and beyond Horndeski  \cite{bh1,bh2} interactions,
\bea
&&\L_2 = G_2(\varphi, X)   \nonumber \\
&&\L_3= G_3(\varphi, X) \Box \varphi  \nonumber \\
&&\L_4 = G_4(\varphi, X)R-2G_{4, X} \varphi_{[\mu_1}{}^{\mu_1}  \varphi_{\mu_2]}{}^{\mu_2}  \nonumber \\
&&  \qquad +F_4(\varphi, X)\epsilon^{\mu\nu\rho}{}_\sigma \epsilon^{\mu'\nu'\rho'\sigma}\varphi_\mu \varphi_{\mu'} \varphi_{\nu}{}_{\nu'} \varphi_\rho{}_{\rho'}   \nonumber \\
&& \L_5 = G_5(\varphi, X) G_{\mu\nu} \varphi^\mu{}^\nu 
 +\frac{G_{5,X}}{3} \varphi_{[\mu_1}{}^{\mu_1} \varphi_{\mu_2}{}^{\mu_2} \varphi_{\mu_3]} {}^{\mu_3}   \qquad \nonumber \\
&&\qquad  + F_5(\varphi, X)\epsilon^{\mu\nu\rho\sigma}\epsilon^{\mu'\nu'\rho'\sigma'}\varphi_\mu  \varphi_{\mu'} \varphi_{\nu}{}_{\nu'} \varphi_\rho {}_{\rho'} \varphi_{\sigma}{}_{\sigma'} \qquad  \nonumber
\eea
\normalsize
where $R$ is the Ricci scalar and $G_{\mu\nu}$ is the  Einstein tensor. The symbol  $\epsilon^{\mu\nu\rho\sigma}$ is the totally antisymmetric Levi-Civita tensor, satisfying $\epsilon^{\mu\nu\rho\sigma}\epsilon_{\mu\nu\rho\sigma}=-24$. The  square brackets denote antisymmetric combinations defined without the usual  factors of $1/n!$.  As is well known, despite the higher order nature of these interactions, (beyond) Horndeski theories do not automatically propagate additional  degrees of freedom beyond the scalar and the metric, and can remain free from Ostrogradski ghosts \cite{ostro}. Indeed, the theory will propagate one scalar and two graviton degrees of freedom in each of the following cases \cite{dhost3}:   the Horndeski class \cite{horn} with second order field equations for which $F_4=F_5=0$; beyond Horndeski \cite{bh1,bh2} with $\L_4=0, G_{5, X} \neq 0$ {\it or} $\L_5=0, G_4-2XG_{4, X} \neq 0$;
    beyond Horndeski  with both  $\L_4 \neq 0, \L_5 \neq 0$ and a degeneracy condition 
   \begin{equation} 
   XG_{5,X}F_4 =3F_5\left[ G_4-2XG_{4,X} -(X/2)G_{5,\varphi}\right]\,.
   \end{equation}
 We now consider fluctuations on a homogeneous and isotropic background with scalar $\varphi=\varphi(t)$, metric, 
\begin{equation}
	\dd s^2 = g_{\mu\nu}d x^\mu \dd x^\nu = -\dd t^2 + a(t)^2\delta_{ij}\dd x^i\dd x^j,
\end{equation}
and a vanishing gauge field $A_\mu=0$. As is well known, the tensor fluctuations for the beyond Horndeski action are given by \cite{kob, bh1, bh2}
\begin{equation}
	S_T^{(2)}[h_{ij}] = \frac{1}{8}\int \dd t\dd^3x\ a^3\left[{\mathcal{G}_T}\dot{h}_{ij}^2 - \frac{\mathcal{F}_T}{a^2}(\partial_kh_{ij})^2\right], \label{Tpert}
\end{equation}
where
\begin{align}
    \mathcal{F}_T&\equiv 2G_4 + XG_{5,\varphi}-2X\ddot{\varphi}G_{5,X},  \\
    \begin{split}
        \mathcal{G}_T&\equiv 2G_4-4XG_{4,X}-XG_{5,\varphi}+2X^2F_4\\
        &\qquad\quad -2HX\dot{\varphi}(G_{5,X}+3XF_5),	
    \end{split}
\end{align}
and $H\equiv \dot{a}/a$, $X=-\dot \varphi^2$. The speed of the gravitational wave through the cosmological medium is thus $c_T^2=\mathcal{F}_T/\mathcal{G}_T$.

To derive the propagation of the electromagnetic wave, we consider fluctuations in the photon field in the Coulomb gauge where $A_0=0$ and $\del^i A_i=0$, such that we obtain 
\begin{equation}
	S_\gamma^{(2)}[A_i] =\frac12 \int \dd t\dd^3x\ a \left[{\mathcal{G}_\gamma}\dot A_{i}^2 - \frac{\mathcal{F}_\gamma}{a^2}(\del_i A_j)^2\right], \label{Flucgam}
\end{equation}
with
\begin{equation}
    \mathcal{F}_\gamma\equiv \lambda\,, \qquad \mathcal{G}_\gamma \equiv	\lambda+\frac{\mu X}{2}\,.
\end{equation}
The electromagnetic wave propagates through the cosmological  medium at speed $c_\gamma^2=\mathcal{F}_\gamma/\mathcal{G}_\gamma$. Clearly this need not be equal to unity. 

In order to satisfy the multi-messenger speed tests, we require that $c_T^2=c^2_\gamma$, or equivalently
\begin{equation}
    \mathcal{F}_T \mathcal{G}_\gamma- \mathcal{F}_\gamma\mathcal{G}_T=0\,.\label{detFG}
\end{equation}
Requiring this to vanish on any cosmological background, we set the coefficients of powers of  $\ddot \varphi$ and $H$ to vanish independently. This yields three equations
\begin{align}
0&=  \mathcal{F}_T|_1\left(\lambda+\frac{\mu X}{2}\right)   -\lambda  \mathcal{G}_T|_1, \\ 
0&= \mathcal{F}_T|_{\ddot \varphi}\left(\lambda+\frac{\mu X}{2}\right), \\
0&= -\lambda \mathcal{G}_T|_H,
\end{align}
where 
\begin{align}
    \mathcal{F}_T|_1&\equiv 2G_4 + XG_{5,\varphi}, \label{Fcoeff1}\\
    \mathcal{F}_T|_{\ddot \varphi} &= -2XG_{5,X}, \label{Fcoeff2} \\
        \mathcal{G}_T|_1&\equiv 2G_4-4XG_{4,X}-XG_{5,\varphi}+2X^2F_4, \label{Fcoeff3}\\
         \mathcal{G}_T|_H&\equiv -2X\sqrt{-X}(G_{5,X}+3XF_5),	\label{Fcoeff4}
\end{align}
and we have used the fact that $\dot \varphi=\sqrt{-X}$. Avoiding the singular limit in which  $ \mathcal{F}_\gamma$ and/or $ \mathcal{G}_\gamma$ vanishes, this implies that $G_5=G_5(\varphi), ~F_5=0$ and 
\begin{equation}
    \frac{\mu}{\lambda}=\frac{4(XF_4-G_5'-2G_{4, X})}{2G_4+XG_5'}, \label{mulam}
\end{equation}
where prime denotes differentiation with respect to $\varphi$. Note that the degeneracy condition is also satisfied.

The constraints presented in \cite{Sakstein:2017xjx,Ezquiaga:2017ekz,Creminelli:2017sry,Baker:2017hug,Amendola:2017orw,Langlois,Marco,boss,Kase}, where they required $c_T^2=1$, are recovered in the limit where $\mu=0$. This fixes $F_4=(G_5'-2G_{4, X})/X$, drastically reducing the parameter space of allowed theories. By allowing a disformal coupling between the photon and the scalar through non-vanishing values of $\mu$, it seems that the parameter space of allowed theories is increased.  Note that for $G_5=F_5=0$, we recover the solution presented in \cite{Mironov:2024idn}, at least for a constant dilaton. 

At this stage, we also impose an orthogonal pheonomological constraint derived in \cite{decay}. The fact that the gravitational wave arrived at all, and did not decay into scalars,  means that the following should be negligible
\begin{equation}
    M^2+2(\tilde m_4^2 -m_4^2)-c_T^2(M^2-2m_5^2)=0, \label{decay}
\end{equation}
where 
\begin{align}
M^2 &= \mathcal{G}_T \nn, \\
% 2 G_4 - 4 X G_{4,X} - X \big( G_{5,\varphi} + 2 H \dot \varphi  G_{5,X}  \big) 
% \nn \\
% & \qquad  + 2 X^2 F_4 -6H \dot \phi X^2 F_5\;,
% \nn \\
m_4^2 & = \tilde m_4^2 + X^2 F_4 - 3 H \dot \varphi X^2 F_5   \;,\nn  \\
\tilde m_4^2 & = - \big[ 2 X G_{4,X} + X G_{5,\varphi} + \big(H \dot \varphi - \ddot \varphi \big) X G_{5,X}  \big] \;, \nn \\
m_5^2 & = 
 X \big[ 2 G_{4,X} +   4 X G_{4,XX} + H \dot \varphi  ( 3 G_{5,X}  + 2  X  G_{5,XX}) \nn \\
&\qquad  + G_{5,\varphi}  + X G_{5,X \varphi}   - 4 X F_4 -2X^2 F_{4,X}   
\nn \\ 
&\qquad + H \dot \varphi  X \big( 15 F_5 + 6 X F_{5,X} \big) \big], \nn
\end{align}
 correspond to specific couplings in the effective theory of dark energy \cite{eftde}. For this to be true independently of $H$ and $\ddot \varphi$, it follows that
\begin{equation}
    F_4=\frac{2H_{4, X}}{X}-\frac{H_4}{X^2}+\frac{J_4(\varphi)}{X^2 H_4}, \label{F4}
\end{equation}
where $H_4=G_4(\varphi, X)+\frac12 XG_5'(\varphi) \neq 0$, $J_4(\varphi)$ is an arbitrary  function and  we have also used the fact that $G_5=G_5(\varphi), ~F_5=0$. Plugging this into \eqref{mulam} we find that we must also have
\begin{equation}
     \frac{\mu}{\lambda}=
     \frac{2}{X}\left(\frac{J_4(\varphi)}{H_4(\varphi,X)^2}-1\right). \label{mulam2}
\end{equation}
Although we appear to have extended the parameter space of allowed theories, we will now show that this is just an artefact of a disformal transformation. To this end, consider a Lagrangian for the electromagnetic field  of the form
\begin{equation}
    \L_\gamma=   - \frac{1}{4}\rho(\varphi,X) F_{\mu\nu}F^{\mu\nu}.  \label{Lgammin}
\end{equation}
As $\mu=0$ it is clear that the electromagnetic wave propagates with unit speed, $c_\gamma^2=1$. We further assume that this Lagrangian sits alongside a beyond Horndeski theory whose gravitational waves propagate at an undetermined speed $c_T^2$ in the cosmological medium. 

We now perform a disformal transformation
\begin{equation}\label{eq:disformal}
		g_{\mu\nu}\mapsto \widetilde{g}_{\mu\nu} = A(\phi,X)g_{\mu\nu} + B(\phi,X)\phi_\mu\phi_\nu\,.
	\end{equation}
Note that for the transformation \eqref{eq:disformal} to be invertible we must have $A(A-A_X X-B_X X^2)\neq 0$. Further, for the new metric to be  non-degenerate we also require $A(A+XB) \neq 0$. Indeed, one can readily show that the inverse metric transforms as
\begin{equation}
	\widetilde{g}^{\mu\nu} = \frac{1}{A}g^{\mu\nu}-\frac{B}{A(A+BX)}\phi^\mu \phi^\nu\equiv\mathcal{A}g^{\mu\nu}+\mathcal{B}\phi^\mu\phi^\nu,
\end{equation}
and the determinant as 
\begin{equation}
	\sqrt{-\widetilde{g}} = A\left(A+BX\right)\sqrt{-g}\,.
\end{equation}

As beyond Horndeski theories are known to be closed under disformal transformations, the gravitational sector remains in the beyond Horndeski class, albeit with a possibly different propagation speed for the gravitational wave, which we label $\tilde c_T^2$. We will compute $\tilde c_T^2$ in a moment. First, let us consider the impact of the disformal transformation on the electromagnetic sector. It turns out that the Lagrangian \eqref{Lgammin} is transformed in to the general form given by \eqref{Lgam}, with functions $\lambda=f\mathcal{A}^2$ and $\mu=2 f\mathcal{AB}$ and $f(\varphi, X)=\rho(\varphi, \widetilde X) \mathcal{A}(\mathcal{A}+\mathcal{B}X)$.  Here the disformal transformation of the kinetic term $\widetilde X=\widetilde g^{\mu\nu} \varphi_\mu \varphi_\nu=\mathcal{A}X+\mathcal{B}X^2 $. There is enough freedom in the disformal transformation for the two functions $\lambda$ and $\mu$ to be completely arbitrary. We immediately see that the electromagnetic wave in the disformally transformed theory propagates at a speed
\begin{equation}\label{eq:chat}
	\tilde {c}_\gamma^2 = \frac{\lambda}{\lambda+\frac{\mu X}{2}} =\frac{A+BX}{A}\,.
\end{equation}
In the limit where the  disformal metric  becomes degenerate, $A(A+BX )\to 0$, note that the speed of the electromagnetic wave either vanishes or diverges.

We now consider the effect of the disformal transformation on the gravitational sector. Recall that we started with a generic beyond Horndeski theory with tensor fluctuations described by \eqref{Tpert} and gravitational wave speed given by $c_T^2=\mathcal{F}_T/\mathcal{G}_T$. The key point is that the  tensor perturbation, $h_{ij}$, is disformally invariant and so it is sufficient to perform the disformal transformation on the background. To this end, we note that
\begin{equation}
    \dd t\to \widetilde N \dd t, \qquad \del_t \to \frac{1}{\widetilde N} \del_t, \qquad a \to \sqrt{A} a,
\end{equation}
where $\widetilde N=1/\sqrt{-\widetilde g^{tt} }=\sqrt{A+B X}$.  It follows that 
\begin{eqnarray}
\mathcal{F}_T(\varphi, \dot \varphi, \ddot \varphi, a , \dot a)   &\to & \widetilde{\mathcal{F}}_T=\sqrt{A} \widetilde N\mathcal{F}_T\left(\varphi, \frac{ \dot \varphi}{\tilde N}, \frac{\ddot \varphi}{\tilde N^2} -\frac{\dot{\tilde N} \dot \varphi}{\tilde N^3}, a , \frac{\dot a}{\tilde N} \right), \nn  \\
\mathcal{G}_T(\varphi, \dot \varphi, \ddot \varphi, a , \dot a) &\to & \widetilde{\mathcal{G}}_T=\frac{A^{3/2}}{\widetilde N}\mathcal{G}_T\left(\varphi, \frac{ \dot \varphi}{\tilde N}, \frac{\ddot \varphi}{\tilde N^2} -\frac{\dot{\tilde N} \dot \varphi}{\tilde N^3}, a , \frac{\dot a}{\tilde N} \right), \nn
\end{eqnarray}
suggesting that the gravitational wave in the transformed system propagates at speed
\begin{equation}
    \tilde c_T^2=\frac{\widetilde N^2}{A} c_T^2=\tilde c_\gamma^2 c_T^2\,. \label{c-cond}
\end{equation}
In the first part of this section, we demanded that the gravitational wave in a generic beyond Horndeski background propagates at the same speed as its electromagnetic counterpart governed by \eqref{Lgam}. We now see that this amounts to setting $\tilde c^2_T=\tilde c_\gamma^2$ in the current context. From  \eqref{c-cond}, this is equivalent to imposing $c_T^2=1$, prior to the disformal transformation. 

It is in this sense that the new theories satisfying \eqref{mulam} for general $\mu$ do not really represent an increase in the parameter space of theories allowed by multi-messenger speed tests. They are nothing more than general disformal transformations of theories that have gravitational waves propagating at unit speed, exposed in  \cite{Sakstein:2017xjx,Ezquiaga:2017ekz,Creminelli:2017sry,Baker:2017hug,Amendola:2017orw,Langlois,Marco,boss,Kase}.

However, as we mentioned in the introduction,  we can break this disformal degeneracy by coupling other sectors minimally to different disformal combinations of the metric. As an example, consider a Fab-Four like theory  \cite{Fab4-1, Fab4-2, Fab4-3}, supported by ``John" and ``George" terms, such that $G_5=F_4=F_5=0$ and $G_4=V_\text{john} (\varphi) X+V_\text{george}(\varphi) $.  The non-trivial ``John" term was originally ruled out by multi-messenger speed tests. However, this is no longer the case provided we include a disformal coupling to the photon with 
$$
\frac{\mu}{\lambda}=-\frac{4V_\text{john}}{V_\text{john} X+V_\text{george}}.
$$
In principle, all other matter fields can now be mininally coupled to the metric in this ``Fab Four" frame, just as they are in the original model.  Further phenomenological constraints are now required to rule out the ``John" couplings. In this case, it turns out that we can use constraints on the decay of gravitational waves to set $V_\text{john}=0$ \cite{decay}. 

Although the Fab Four are unable to pass all the constraints, this is not the case with the so-called beyond Fab Four theory \cite{bF4}. This is also designed to self-tune the cosmological constant but includes beyond Horndeski terms. In particular, we see that $V_\text{john}(\varphi)\to F_\text{john}(\varphi, X)$, such that  $G_4 \to  -\frac12 X F_\text{john}(\varphi, X) + V_\text{george}(\varphi)$, $F_4 \to  F_\text{john,X}$. In this case, matching speeds requires that
\begin{equation}
    \frac{\mu}{\lambda}=\frac{4(2XF_\text{john,X}+F_\text{john})}{2V_\text{george}- XF_\text{john}}, \label{mulambF$}
\end{equation}
while the decay constraint \eqref{F4} amounts to a non-linear differential equation for $F_\text{john}$.

The lesson to take away is that multi-messenger speed tests can open up the parameter space of possibilities up to a disformal transformation. This can alter the status of some modified gravity theories in a positive way. Other orthogonal constraints are then required to rule them out, if at all.

\section{More general scalar-photon couplings} \label{sec:gen-couplings}

Although the photon Lagrangians \eqref{Lgam} inspired by \cite{Mironov:2024idn} amount to a $U(1)$ gauge field minimally coupled to a disformal metric,  other interactions are possible at leading order in the electromagnetic field strength \cite{Heisenberg:2018acv,Heisenberg:2018mxx,Kase:2018nwt}. Despite being higher order in derivatives, these interactions do not propagate any additional degrees of freedom, nor can they be mapped to the minimal set-up via an invertible field redefinition.

Working up to quadratic order in the field strength, a more general Lagrangian for the electromagnetic sector is given by 
\begin{equation}
    \L_\gamma=\mathcal{L}_{\text {SVT}}^{(2)}+\mathcal{L}_{\text {SVT}}^{(3)}+\mathcal{L}_{\text {SVT}}^{(4)},
\end{equation}
where
\begin{align} \label{eq:L_SVT}
   \mathcal{L}_{\text {SVT}}^{(2)}&=  - \frac{1}{4}\left[\lambda(\varphi,X) F_{\mu\nu}F^{\mu\nu} + \mu(\varphi,X)F^{\mu}{}_{\alpha}{F}^{\nu\alpha}\varphi_\mu\varphi_\nu\right], \\
\mathcal{L}_{\text {SVT}}^{(3)} &= -\frac14 \left[ f_3(\varphi, X) g_{\alpha \beta}+\tilde{f}_3(\varphi,X) \varphi_\alpha \varphi_\beta
\right] \tilde{F}^{\mu \alpha} \tilde{F}^{\nu \beta} \varphi_{\mu\nu},
\\
\mathcal{L}_{\mathrm{SVT}}^4&=-\frac14 \left[ f_4(\varphi, X) P^{\mu \nu \alpha \beta} F_{\mu \nu} F_{\alpha \beta} \right. \nn \\ 
&\qquad \qquad \left.-4f_{4, X} \tilde{F}^{\mu \nu} \tilde{F}^{\alpha \beta} \varphi_{\alpha\mu} \varphi_{\beta\nu} \right],
\end{align}
where the dual of electromagnetic field strength, 
$
\tilde{F}^{\mu \nu}=\frac{1}{2} \epsilon^{\mu \nu \alpha\beta} F_{\alpha \beta},
$ and the double dual of the Riemann tensor $P^{\alpha \beta \gamma \delta}=\frac{1}{4} \epsilon^{\alpha \beta \mu\nu} R_{\mu \nu \rho \sigma} \epsilon^{\rho \sigma \gamma \delta}$.

Although the structure of  $\mathcal{L}_{\text {SVT}}^{(2)}$ is familiar from \cite{Mironov:2024idn} and the previous section, the remaining terms are qualitatively new. Let us now explore their impact on the speed of propagation of the electromagnetic wave.  Fluctuations of the photon field in Coulomb gauge take the form \eqref{Flucgam}, with
\begin{align}
& \mathcal{F}_\gamma=\lambda+
\frac12 H \dot \varphi\left( f_3+\tilde f_3 X \right)+\frac{f_3}{2}\ddot{\varphi}
  \nn \\
  & \qquad \qquad 
  +4 f_{4, X} H \dot{\varphi} \ddot{\varphi}-2 f_4 \left(\dot{H}+H^2\right), \\
& \mathcal{G}_\gamma=\lambda+\frac{\mu}{2} X+H f_3 \dot{\varphi}
-2H^2(f_4+2X f_{4, X}),
\end{align}
and  a propagation speed $c_\gamma^2 = \mathcal{F}_\gamma/\mathcal{G}_\gamma$. Requiring this to agree with the speed of the gravitational wave, we impose \eqref{detFG}. This is now a more complex equation containing powers of $\ddot \varphi$, $H$ and $\dot H$ with coefficients that depend on $\varphi$ and $X$. We now set those coefficients to vanish on an arbitrary cosmological background. This is a conservative approach. In principle, we could use the background equations of motion to relate $\ddot \varphi$, $H$ and $\dot H$ to just the energy density and pressure and reduce the number of constraints. In any event, avoiding singular limits as before, the coefficient of $\dot H$ immediately constrains $f_4=0$. The vanishing of the other coefficients yields
\begin{align}
0&=  \mathcal{F}_T|_1\left(\lambda+\frac{\mu X}{2}\right)   -\lambda  \mathcal{G}_T|_1 ,\label{newcon1st}\\ 
0&= \mathcal{F}_T|_{\ddot \varphi}\left(\lambda+\frac{\mu X}{2}\right) - \mathcal{F}_\gamma|_{\ddot \varphi}\mathcal{G}_T|_1, \label{newconddp1}\\
0&= \mathcal{F}_T|_1\mathcal{G}_\gamma|_H-\lambda \mathcal{G}_T|_H-\mathcal{F}_\gamma|_H \mathcal{G}_T|_1, \label{newconH1}\\
0&= -\mathcal{F}_\gamma|_H \mathcal{G}_T|_H, \label{newconHH} \\
0&= \mathcal{F}_T|_{\ddot \varphi} \mathcal{G}_\gamma|_H-\mathcal{F}_\gamma|_{\ddot \varphi} \mathcal{G}_T|_H, \label{newconlast}
\end{align}
where we recall \eqref{Fcoeff1} to \eqref{Fcoeff4} and further define 
\begin{align}
 \mathcal{F}_\gamma|_H&=
\frac12 \sqrt{-X} \left( f_3+\tilde f_3 X \right), \label{FgH}\\
\mathcal{F}_\gamma|_{\ddot \varphi} &=
\frac{f_3}{2}, \label{Fgddp} \\
 \mathcal{G}_\gamma|_H &=f_3 \sqrt{-X}. \label{GgH}
\end{align}
Before solving these constraints directly it is convenient to first use the orthogonal constraint derived in \cite{decay}.  The fact that the gravitational wave  did not completely decay into scalars  means that the combination of terms given by \eqref{decay} should be negligible.  When we impose this alongside the constraints \eqref{newcon1st} and \eqref{newconlast}, the only non-singular solutions have $f_3=\tilde f_3=0$, explored in detail the previous section. We conclude that the higher order couplings  revealed in \cite{Heisenberg:2018acv,Heisenberg:2018mxx,Kase:2018nwt} are ruled out by a combination of multi-messenger speed tests and decay constraints.  The only non-trivial scenarios are those which are disformally related to modified gravity theories with unit speed of gravitational waves, also satisfying \eqref{F4} and \eqref{mulam2}. A detailed derivation of this result is found in the appendix.

\section{Discussion}\label{sec:conc}
In this paper we have revisited the impact of multi-messenger speed tests on the parameter space of allowed theories of modified gravity, falling within the class of Horndeski and beyond Horndeski theories.  This was motivated by recent work  \cite{Mironov:2024idn}  in which  Kaluza-Klein compactifications generate non-trivial couplings between the dark energy scalar and a $U(1)$ gauge field, and are then shown to be compatible with experimental constraints. Although the gravitational wave propagates at non-unit speed on the cosmological background, the same is true of the electromagnetic wave, and the two speeds can still coincide. Here we have shown that these kind of scalar-photon couplings are equivalent to an electromagnetic field minimally coupled to a disformal metric. This coupling can then be undone by a disformal transformation, such that we recover the original constraints on modified gravity theories, up to and including couplings between the gravitational and electromagnetic fields and dark energy. Of course, this disformal equivalence can be broken by the gravitational couplings to other sectors. 

We also considered a more general class of scalar-photon couplings that cannot be undone by disformal transformations. Once again we find scenarios in which both the gravitational and electromagnetic waves can propagate at the same speed,  different from unity. However, when we impose constraints on the decay  of gravitational waves, these additional scenarios are ruled out.  We are left with the class of modified gravity disformally related to those with unit gravitational wave speed, also satisfying \eqref{F4} and \eqref{mulam2}.

Of course, non-minimal couplings between the gravity, electromagnetism and dark energy will be subject to other phenomenological constraints. For example, suppose the action is analytic around $\varphi=0$ and we can a perform Taylor expansion,  $\lambda(\varphi, X)=1+\mathcal{O}(\varphi, X)$ and $\mu(\varphi, X)=\frac{1}{\Lambda^4} +\mathcal{O}(\varphi, X)$. This generates a leading order scalar-photon interaction of the form $\frac{1}{\Lambda^4}F^\mu{}_\alpha F^{\nu \alpha}\varphi_\mu \varphi_\nu$. Assuming  canonical kinetic terms for all fields at leading order, this coupling is constrained to be $\Lambda \gtrsim 400$ GeV, through mono-photon searches \cite{Brax_2014, Brax:2015hma}. 
To the best of our knowledge, direct constraints on the other disformal interactions present in \eqref{Lgam} have not been computed. 

In a realistic set-up, our theory will also be coupled to other standard model fields.  Couplings between the dark energy field and Standard Model fermions and gauge bosons are strongly constrained by astrophysical and laboratory tests \cite{Brax_2014, Brax:2015hma, Sakstein:2014isa, Brax:2012ie, Brax:2015fya}. Interestingly, the dark energy coupling to neutrinos can also be tested through the impact on their dispersion relation. For example, this could be through the coincident emission of gravitational waves and neutrinos from mergers of super-massive black holes, or a neutron star and a stellar-mass black hole. Such gravitational waves would be detectable with mHz interferometers (e.g LISA), and the neutrino radiation through ground-based experiments (e.g IceCube).  Another similar possibility of coincident GW-neutrino radiation is through observations of explosive events such as gamma-ray bursts\footnote{For a review on such multi-messenger probes of various astrophysical systems see \cite{Murase_2019}.}. 
The detectability of these effects will clearly depend on the size of the effect and the observational/theoretical uncertainties (neutrino masses, astrophysical uncertainties etc). Interestingly, a coincident detection of photons and neutrinos from a flaming blazar has been recently observed \cite{flaring_blazar}.

Yet another way to test the scenario described in this work is through structure formation.  Whenever $c_T=c_\gamma \neq 1$,  structure formation will no longer align with the predictions of $c_T = 1$ models. In \cite{Amendola:2017orw} it was shown that the luminality of gravitational waves implies that the growth of matter at large scales can be at least as fast as in GR. However, allowing for $c_T \neq 1$ implies that matter can now cluster more strongly or more weakly at large scales, depending on the choice of theory space. Although coincident GW and electromagnetic radiation will no longer constrain the theory, the propagation of gravitational waves will depart from GR: their amplitude will be damped due to the running of the effective gravitational strength, which is qualitatively similar to the case of conformally-coupled theories. However, $c_T \neq 1$ will also lead to a modified evolution of the wave's phase compared to GR \cite{LISA_2022}, a distinct signature of the graviton-scalar derivative interactions.

Finally, it is important to emphasize that the analysis presented here does not make any use of the background equations of motion.  Putting the multi-messenger speed constraints on shell reduces the number of conditions, as in \cite{Copeland:2018yuh}. It would be interesting to revisit our work in the same spirit. However, we also note that inhomogeneities are expected  to close off any additional loopholes that could arise, as was the case in \cite{Bordin:2020fww}. We have also focused on Horndeski and beyond Horndeski theories, which are closed under disformal transformatons.  It would be interesting to extend the analysis to the so-called DHOST theories \cite{dhost1,dhost2,dhost3} which also describe the dynamics of a massless graviton and a single scalar degree of freedom. The sight and sound of  neutron star mergers  have been a powerful tool in constraining modified theory of gravity in the late universe. However, our analysis makes it abundantly clear that multi-messenger tests are not just sensitive to how the dark energy field couples to gravity, but also how it couples to electromagnetism  and other forms of matter.

\section*{Acknowledgments}
CC and AP thank CEICO/FZU for their hospitality when this work was initiated. EB acknowledges support of ANR grant StronG (ANR-22-CE31-0015-01).  CC acknowledges support by the French National Research Agency via Grant No. ANR-20-CE47-0001 associated with the project COSQUA (Cosmology and Quantum Simulation).  AP acknowledges partial support from the STFC Consolidated Grant nos. ST/V005596/1, ST/T000732/1, and ST/X000672/1. IDS acknowledges funding from the Czech Grant Agency (GA\^CR) under the grant number 21-16583M. For the purpose of open access, the authors have applied a CC BY public copyright licence to any Author Accepted Manuscript version arising.

\appendix

\section{Can $f_3$ or $\tilde f_3$ be non-vanishing?} \label{sec:App}
In this appendix, we prove that there are no consistent scenarios with $f_3$ or $\tilde f_3$ non-zero, for which we can satisfy the decay constraint \eqref{decay} and matching speeds, $c_T^2=c_\gamma^2 $.

We have already seen how the $\dot H$ dependence in the speed matching  requires $f_4=0$. We proceed by assuming that at least one of  $f_3$ or $\tilde f_3$ is non-zero.  When $c_T^2=c_\gamma^2$, the the decay constraint \eqref{decay} is  equivalent to $\mathcal{G}_i Y=\mathcal{F}_iZ$ for $i=T, \gamma$ and
\begin{equation}
Y=    M^2+2(\tilde m_4^2 -m_4^2), \qquad Z=M^2-2m_5^2. \label{decayapp}
\end{equation}
Observe that $Y, Z$ and $\mathcal{G}_i$ are independent of $\ddot \varphi$. It follows that $\mathcal{F}_i|_{\ddot \varphi}Z=0$.

First consider the case where $Z \neq 0$. It follows that $\mathcal{F}_i|_{\ddot \varphi}=0$, giving $f_3=0$ and $G_5=G_5(\varphi)$. As a result, we now require that $\tilde f_3 \neq 0$.  As far as the speed tests are concerned,  \eqref{newconddp1} and \eqref{newconlast} hold automatically.

We now apply \eqref{newconHH}: since $f|_3=0, \tilde f_3 \neq 0$ we note from \eqref{FgH} that $\mathcal{F}_\gamma|_{H} \neq 0$ and so we infer that $\mathcal{G}_T|_{H}=0$.  Since $G_5=G_5(\varphi)$, it follows that $F_5=0$. 

We now apply \eqref{newconH1}: when $f_3=f_4=F_5=0$ and $G_5=G_5(\varphi)$, it follows that $\mathcal{G}_i|_H=0$ and we are left with $\mathcal{F}_\gamma|_{H} \mathcal{G}_T|_1 =0$. Since we have already noted that $\mathcal{F}_\gamma|_{H} \neq 0$, we must have $\mathcal{G}_T|_1 =0$.  Unfortunately, we now have  $\mathcal{G}_T|_1 = \mathcal{G}_T|_H =0$ which means we are in the singular limit $\mathcal{G}_T=0$. 

Next we consider the other case where  $Z = 0$. To be consistent with $\mathcal{G}_i Y=\mathcal{F}_iZ$ away from the singularities, we must also have $Y=0$. We  can write $Q=Q|_1(\varphi, X)+HQ|_H(\varphi, X)$ for $Q=Y, Z$, requiring them to vanish independently of $H$. We obtain four equations $Y|_1=Y|_H=Z|_1=Z|_H=0$. These can be integrated directly to give $G_5=G_5(\varphi), F_5=J_5(\varphi)/X^3, G_4=-\frac12 XG_5' +\sqrt{-X}J_4(\varphi)$ and $F_4=J_4(\varphi)/\sqrt{-X} X^2$.

Let us now turn to the speed constraints. First up, \eqref{newconlast} gives $\mathcal{F}_\gamma|_{\ddot\varphi}\mathcal{G}_T|_H=0$ since $\mathcal{F}_T|_{\ddot\varphi}$ vanishes when $G_5=G_5(\varphi)$. From \eqref{newconHH} we also have $\mathcal{F}_\gamma|_{H}\mathcal{G}_T|_H=0$. Note that these two equations are equivalent to $f_3  \mathcal{G}_T|_H=\tilde f_3\mathcal{G}_T|_H=0$. Since $f_3$ and $\tilde f_3$ cannot vanish simultaneously by assumption, we must have $\mathcal{G}_T|_H=0$, which implies that $J_5=0$ and so $F_5=0$.

Now consider \eqref{newconddp1} with $\mathcal{F}_T|_{\ddot \varphi}=0$.We further assume that $\mathcal{G}_\gamma|_1 \neq 0$ in order to avoid the singular limit. It follows that $\mathcal{F}_\gamma|_{\ddot \varphi}=0$ , or equivalently, $f_3=0$. This also implies that  $\mathcal{G}_\gamma|_H=0$.  Finally, we consider \eqref{newconH1}: using the fact that $\mathcal{G}_\gamma|_H=\mathcal{G}_T|_H=0$, we obtain $\mathcal{F}_\gamma|_H=\mathcal{G}_T|_1=0$. This is  problematic: $\mathcal{F}_\gamma|_H=0$ would imply that we now have $f_3=\tilde f_3=0$,  violating our opening assumption; $\mathcal{G}_T|_1=0$, along with the fact that $\mathcal{G}_T|_H=0$ would imply that we are in the singular limit $\mathcal{G}_T=0$.

We conclude that there are no consistent scenarios in which either of $f_3$ or $\tilde f_3$ is non-vanishing.

% 
%%%%%%%%%%%%%%%%%%%%%%%%%%%%%%%%%%%%%%

\end{document}